# Evaluation of strategies for efficient rate-distortion NeRF streaming

Pedro Martin, António Rodrigues, João Ascenso, and Paula Queluz
Instituto de Telecomunicações - Instituto Superior Técnico, Lisbon, Portugal
{pedro.martin, antonio.rodrigues, joao.ascenso, paula.queluz}@lx.it.pt

*Abstract*—**Neural Radiance Fields (NeRF) have revolutionized the field of 3D visual representation by enabling highly realistic and detailed scene reconstructions from a sparse set of images. NeRF uses a volumetric functional representation that maps 3D points to their corresponding colors and opacities, allowing for photorealistic view synthesis from arbitrary viewpoints. Despite its advancements, the efficient streaming of NeRF content remains a significant challenge due to the large amount of data involved. This paper investigates the rate-distortion performance of two NeRF streaming strategies: pixel-based and neural network (NN) parameter-based streaming. While in the former, images are coded and then transmitted throughout the network, in the latter, the respective NeRF model parameters are coded and transmitted instead. This work also highlights the trade-offs in complexity and performance, demonstrating that the NN parameter-based strategy generally offers superior efficiency, making it suitable for one-to-many streaming scenarios.**

## I. INTRODUCTION

Recent advancements in 3D visual representations have enabled more immersive and interactive experiences, boosting applications such as autonomous driving, virtual tours, remote education, and gaming. Multiview+depth (MVD), point clouds (PC) and light fields (LF) have become popular in many applications. Nonetheless, these techniques often rely on assumptions that can limit the realism of the 3D representation, particularly for applications where high-fidelity view synthesis is essential.

Neural Radiance Fields (NeRF) have revolutionized the 3D visual representation field, providing unmatched capabilities. They allow realistic view synthesis from a limited set of images achieving high levels of fidelity and detail [1]. Many applications, including 3D reconstruction and virtual reality (VR), may greatly benefit from NeRF technology [2]. NeRF is a volumetric functional representation that is obtained from a set of 2D perspectives of the visual scene. A training process is performed using the spatial location and viewing direction of each 2D perspective, allowing the model to learn a mapping from 3D points to their corresponding colors and opacities. With NeRF, a direct interpolation of the original 2D views is not performed as in classical depth-based image rendering methods. Instead, view synthesis is achieved by predicting color and opacity values for any given 3D point and viewing direction, followed by the application of a classical volume rendering technique. There are two strategies to implement NeRF methods: one with neural networks (NNs) and another without them [2]. The NN-based NeRF employs multi-layer perceptrons (MLPs) to model the radiance field of the scene, leveraging the expressive power of NNs to capture fine details and complex scene variations. On the other hand, NN-free and some hybrid methods rely on directly optimizing voxel grids or other explicit representations. This approach aims to reduce training and rendering time, albeit often at the expense of quality.

A key challenge to enable NeRF driven applications is to find the most efficient way to not only represent a NeRF, but also to transmit it without compromising quality. Therefore, efficient streaming strategies for NeRF-content are needed to handle the large amount of data involved, especially in bandwidth-constrained environments. Several works have already been proposed in the field of NeRF streaming, especially to enable real-time rendering. These approaches focus mainly on methods to compress data to reduce memory consumption, and to improve rendering efficiency [3]-[7]. However, there is no discussion on end-to-end (from capturing to consumption) architectures to transmit NeRF multimedia content considering bandwidth and complexity constraints.

Therefore, the objective of this paper is to examine the streaming performance (regarding rate and distortion) of two distinct NeRF streaming strategies, that differ on the type of data that is coded and transmitted. While in the first (pixel-based) strategy, images are coded, in the second strategy the NN parameters (weights and biases) are coded instead. The pixel-based strategy requires performing the NeRF training locally on the user equipment (UE), utilizing substantial computational resources and a powerful GPU/NPU; thus, it is more suitable for one-to-one streaming applications (such as private cloud storage of NeRF content). The NN parameter-based strategy offloads the NeRF training to a cloud or edge server. This approach is well-suited for one-to-many streaming scenarios, allowing multiple clients to interact with the same content while placing minimal computational demands on their devices. Thus, the main contribution lies in the rate-distortion (RD) performance study of these two strategies. The RD evaluation allows to draw insightful conclusions regarding the strengths and weaknesses of both NeRF streaming solutions, considering the type of the visual scene (e.g., synthetic or real).

The rest of this paper is organized as follows. Section II describes the related work while Section III presents the NeRF fundamental concepts and the selected NeRF method. Section IV describes the NeRF streaming strategies. Section V presents the test sequences and conditions used for evaluation and the experimental results. Lastly, Section VI concludes the paper with some final remarks.

## II. RELATED WORK

NeRF methods that use NN-based (or implicit) visual scene representations [1], [8] have been found to be computationally complex for real-time rendering. Thus, other alternatives were proposed to address this problem using just voxel grids [9], [10], or feature grids and small MLPs (hybrid) [11], [12]. These methods allow to reach real-time rendering but often struggle to perform high-quality synthesis (especially of unbounded large-scale scenes) and consume large amounts of graphics memory.

Therefore, the (so-called) "baking" solutions [3], [4] have emerged as an alternative, converting a trained NeRF to a sparse neural radiance grid by mapping the learned scene features (along with color, volume density) into this grid. In [3], precomputed values from the sparse grid are retrieved and a lightweight MLP is used to evaluate view-dependent effects once per pixel. In [4], a combination of a low-resolution 3D voxel grid and high-resolution 2D feature planes are used. In both cases, the rendering process is significantly accelerated, and memory consumption reduced.

Other related works focus on the compression of the NeRF parameters [5]-[7]. In [5], the vector quantized radiance fields (VQRF) method is proposed to compress volumetric grids used in radiance fields. VQRF employs voxel pruning, vector and weight quantization, and entropy coding. In [6], a compact representation method for grid-based neural fields is introduced, leveraging a novel masking strategy and multi-level wavelet transform. In [7], an end-to-end NeRF framework (named NeRFCodec) is proposed which allows to compress feature planes in a hybrid NeRF representation. The NeRFCodec is based on a learning-based image codec (trained with 2D regular images) and the encoder and decoder heads (i.e., the first and last layers) are adjusted for every visual scene. Although these methods allow to save GPU memory, rendering speed, and the memory cost of the NeRF representation, they do not leverage currently standardized image, video or neural codecs for an efficient transmission. Furthermore, methods [5]-[7] have primarily aimed to reduce memory overhead, without addressing the typical trade-off between bitrate and visual distortion. This paper seeks to fill this gap by proposing and evaluating two NeRF streaming strategies that allow for precise control over the bitrate of the visual scene.

## III. NeRF Creation

This section outlines the fundamentals of NeRF and introduces the method selected for this work, Mip-NeRF 360 [8]. Although various NeRF techniques could be employed, Mip-NeRF 360 delivers superior performance, particularly in complex, real-world 360º visual scenes [13], which is critical for streaming immersive applications. While alternative NeRF methods are feasible, some of them with more compact representations, they often result in lower-quality synthesized images, making Mip-NeRF 360 the preferred choice for achieving the high visual fidelity required in many streaming applications.

### A. NeRF Fundamentals

According to the seminal work of [1], NeRF represents a visual scene by a continuous 5D function, $F_\theta$, that encodes the correspondence between the spatial location of a 3D scene point, $x = (x, y, z)$, for a given viewing direction, $d = (\theta, \varphi)$, and their respective color, $c = (R, G, B)$, and opacity values, $\sigma$, (cf. Fig. 1) [1]. Each process is described in more detail as follows:

*1) NeRF training:* The training main role is to learn the mapping function $F_\theta$ of a given visual scene. This process receives as input a set of training images along with the respective camera poses. The camera poses are either directly obtained (as in the case of synthetic content) or estimated by a structure-from-motion algorithm, such as COLMAP [14], [15] (as in the case of real content scenes). A MLP is used as function $F_\theta$. The NeRF training aims to obtain the MLP parameters by minimizing a loss function which corresponds to the mean squared error (MSE) between the training and the corresponding synthesized images (for the same camera pose of each training image). Due to its nature, the NeRF training process requires high computational capabilities and specialized hardware, as graphics processing units (GPUs).

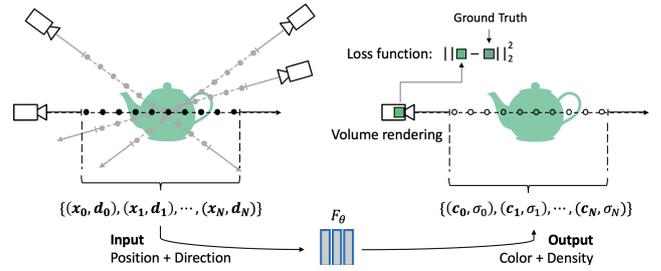

Fig.1. Illustration of the NeRF concept.

*2) View synthesis:* The view synthesis involves querying the learned mapping function $F_\theta$ for multiple points to render the corresponding 2D image. This process takes as input a set of camera poses (trajectory), one for each frame of a video sequence. First, a camera ray is traced for each pixel of the image to be synthesized based on te camera pose. Second, a set of 3D points is sampled along each traced rays. Third, the NeRF model is queried for each sampled point. Finally, to obtain the pixel values corresponding to each camera ray, a classic volume rendering technique is applied to the color and opacity values of the queried points.

### B. Mip-NeRF 360 Method

The Mip-NeRF 360 [8] is a NeRF-based method that reaches state-of-the-art synthesized quality for unbounded (i.e., with foreground and background content) scenes [2]. It is specifically designed for 360º scenes, where the camera moves around a given region of interest, but also shows high performance for front-facing (FF) scenes [13]. In fact, Mip-NeRF 360 is an extension of Mip-NeRF [16]. Mip-NeRF proposed integrated positional encoding and conical frustums for sampling along each pixel ray (instead of 3D points) to reduce aliasing effects and improve the representation of fine details.

Mip-NeRF 360 incorporates several key innovations: *i)* a non-linear scene reparameterization that maps 3D scene coordinates into a bounded domain, effectively managing large-scale unbounded scenes; *ii)* a distortion-based regularizer in the loss function that reduces scene geometry ambiguities (or floaters) and blurriness; and *iii)* a two-MLP architecture, consisting of a small proposal MLP to predict volumetric density, i.e., where in the scene the density (or content) exists, and a large NeRF MLP that uses these locations to make detailed color predictions. This two-MLP architecture enhances the model efficiency, enabling it to handle larger and more complex scenes.

## IV. NeRF Streaming Strategies

The Mip-NeRF 360 training demands significant computational power and energy resources, making its deployment location crucial for supporting applications that use this model. Two potential approaches are proposed: the first involves performing both the Mip-NeRF 360 training and view synthesis directly on the user's device, while the second involves offloading the training to a remote server, such as a cloud or edge server, with view synthesis still occurring on the user's device. In both strategies, view synthesis and rendering must be executed in real-time, responding to user interactions and navigation within the visual scene. An additional scenario, where both training and rendering are handled remotely by the cloud or edge server based on pose information from the user's device, was considered but excluded due to impracticality. This approach involves transmitting pose data from the device to the cloud, where view synthesis and rendering occur, followed by the transmission of images back to the user. However, current network latency and processing time for view synthesis still prevents this cloud streaming scenario to be deployed in practice.

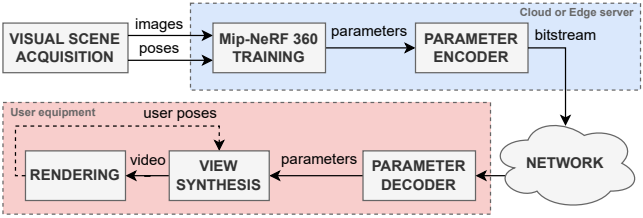

Fig. 2. NN parameter-based streaming strategy pipeline.

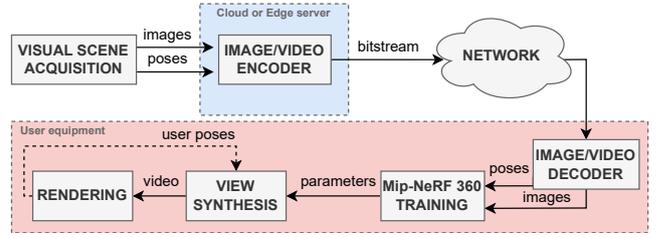

Fig. 3. Pixel-based streaming strategy pipeline.

*A. NN parameter-based streaming strategy*

In this strategy, called as NN parameter-based streaming, the Mip-NeRF 360 training is performed by a cloud server due to its parallel processing capabilities [17]. An edge server (equipped with a powerful GPU/TPU) is also a reliable option due to its quick access and low-latency transmission to end-users [18]. In this approach, the UE is responsible for data acquisition and may also perform view synthesis (including the NN inference). Therefore, devices with limited computational and energy resources, such as smartphones or tablets, can be used. This strategy is suitable for one-to-many (broadcast or multicast) streaming, where the NN parameter representation of the visual scene on the server allows multiple clients (i.e., UEs) to interact with the same content in several ways, enabling applications such as 3D content broadcasting.

This streaming strategy involves transmitting the MLP parameters resulting from the Mip-NeRF 360 training (see Fig. 2). For this purpose, the neural network coding (NNC) standard, which is also known as MPEG-7 part 17, is used [19]. The NNC codec is chosen for its high compression efficiency and versatility for several types of NNs.

The pipeline of this strategy is described as follows:

*1) Visual scene acquisition:* a real camera acquires images or videos of a visual scene. In the case of synthetic content, virtual cameras are used. A selection of acquired images (or frames) is uploaded (after near to visually lossless compression) to the server and the Mip-NeRF 360 is trained with this data. The image selection is usually based on visual coverage of the scene or some type of temporal subsampling (in the case of video acquisition). The training poses can be uploaded along with the training images (if already known) or estimated on the server (otherwise). In the former case, the pose information is transmitted as metadata within the image bitstream.

*2) Mip-NeRF 360 training:* Mip-NeRF 360 is trained on the server with the received training images and respective poses.

*3) Parameter encoder:* The parameters for both MLPs in Mip-NeRF 360 are separately compressed using the NNC codec. The NNC compression process applies a vector quantization scheme, called as dependent scalar quantization, to the MLP parameters, reducing their precision. After, the quantized values are entropy coded using deep context-adaptive binary arithmetic coding (DeepCABAC). DeepCABAC is designed to compress the deep NN parameters efficiently, having achieved high compression ratios across several network architectures and datasets [20]. The reference software of the NNC codec (NNCodec) [21] is used to perform the NNC processes.

*4) Parameter decoder:* The NNC decoder is located in the UE, and the Mip-NeRF 360 parameters are lossy reconstructed from the received bitstream. These reconstructed parameters are an approximation of the original ones and therefore contain some quantization noise.

*5) View synthesis:* The view synthesis process occurs on the UE (allowing a real-time rendering) and uses the decoded Mip-NeRF 360 parameters and the user poses. The quality of the synthesis is conditioned by the compression level of the parameter encoder.

*6) Rendering:* The user poses for view synthesis are calculated according to the user behavior, and the synthesized video is displayed.

*B. Pixel-based streaming strategy*

In this strategy, called as pixel-based streaming, the Mip-NeRF 360 training is performed locally on the UE. This device must possess substantial computational resources and a powerful GPU/NPU, such as those found in a modern laptop. This strategy offers greater flexibility than the previous one since the UE can use any NeRF method. However, it requires more resources since every UE must perform training, which is not particularly advantageous for one-to-many streaming cases. However, it is adequate to one-to-one (or unicast) streaming, enabling applications such as private cloud storage of a gallery of NeRF visual scenes.

This approach consists in the direct transmission of the acquired visual scene (set of images) to the UE, or via a cloud/edge server (see Fig. 3). In case of direct transmission, the acquisition device may also perform image/video encoding. The high efficiency video coding (HEVC) standard, also known as H.265, was selected for that purpose [22]. This codec has high compression efficiency and supports a wide range of resolutions, from low-definition to ultra-high-definition 8K video, and is optimized for several multimedia applications, including streaming, broadcasting, and storage. The key features of HEVC include advanced motion

TABLE I. QP VALUES FOR NNC AND HEVC CODECS FOR EACH VISUAL SCENE.

| Scene | Mugs | | | Antique | | | Train | | | Playground | | |
|---|---|---|---|---|---|---|---|---|---|---|---|---|
| Codec | NNC | HEVC Intra | HEVC Inter | NNC | HEVC Intra | HEVC Inter | NNC | HEVC Intra | HEVC Inter | NNC | HEVC Intra | HEVC Inter |
| QP | -28 | 25 | 18 | -24 | 28 | 21 | -28 | 36 | 28 | -24 | 35 | 27 |
|  | -24 | 30 | 25 | -20 | 35 | 28 | -24 | 40 | 32 | -19 | 39 | 31 |
|  | -20 | 39 | 30 | -16 | 39 | 35 | -20 | 43 | 35 | -15 | 47 | 39 |
|  | -16 | 51 | 51 | -12 | 51 | 51 | -16 | 51 | 51 | -8 | 51 | 51 |

compensation, improved prediction modes, and more flexible block structures.

The new (not described in the previous Section) components of this NeRF streaming strategy pipeline are described next:

*1) Image/video encoder:* The selected images uploaded to the server (see description on previous Section) are compressed by the HEVC encoder. The codec can compress images individually or join them together in a video sequence (as pre-processing) to exploit the temporal redundancy. Therefore, there are two alternative coding modes: intra-frame and inter-frame. The HEVC reference software HM16.2+SCC-8.8 with screen content coding (SCC) [23] was used. The all intra (AI) configuration is used for the intra-frame mode, and the random access (RA) for the inter-frame mode. In both configuration files, the internal bit depth is set to 10 bits, the chroma quantization parameter offsets to 6, and the intra-block copy search area to 3 coding tree units (CTUs) as in [24].

*2) Image/video decoder:* The HEVC decoder is located on the UE and reconstructs a lossy version of the training images. These decoded images have some loss of fidelity, but they are not visualized by the final user and thus may have lower quality, as long as the final rendered images have the desired quality.

Finally, it is important to highlight that the Mip-NeRF 360 method is trained with the decoded training images and not the acquired images as in the previous strategy.

## V. PERFORMANCE EVALUATION

This section presents the test conditions for the performance evaluation of the streaming strategies and the corresponding results and analysis.

### A. Test Sequences

Four visual scenes have been selected from the Neural View Synthesis Quality Assessment (NVS-QA) database [13], namely *mugs* (from the front-facing synthetic scene class), *antique* (from the FF real scene class), *train* and *playground* (from the 360º real scene class). Real content is captured with an optical sensor camera system. In 360º scenes, the camera moves around a specific region of interest with a circular inward-pointing movement. FF scenes involve the camera pointing toward the scene, covering a region of interest, and restricting its motion to a vertical plane. The number of training images of each visual scene is 300, 251, 258, and 275 images, for sequences *mugs*, *antique*, *train*, and *playground*, respectively, all with 960×536 pixels. The frame rate used to capture each visual scene is 25 fps for *mugs* and 30 fps for the remaining, and the synthesized video duration is 10s (leading to 250 and 300 rendered images, respectively).

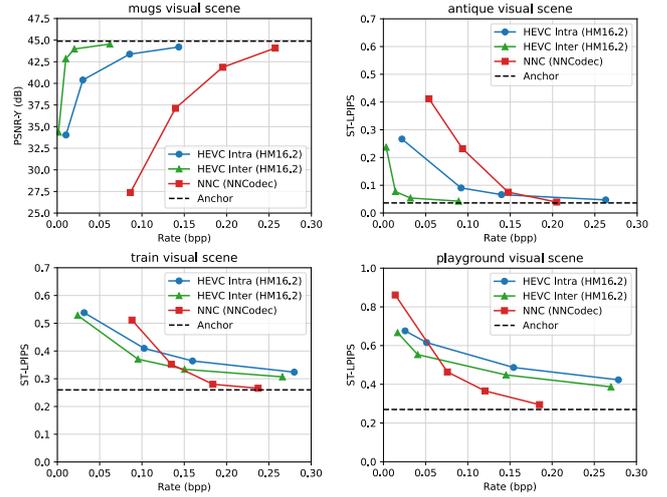

Fig. 4. RD performances of the NeRF streaming strategies for each scene.

### B. Test Conditions

The Mip-NeRF 360 code provided by the authors in [25] was used for training and synthesis. The training was performed with two NVIDIA GeForce RTX 4090 GPUs. The number of iterations for the training process was set to 400 k, a good compromise between training time and synthesis quality. The average training time was 24 hours. Table I shows the NNC and HEVC quantization parameters (QP) used in the two streaming strategy encodings. The streaming strategies performances are compared in terms of RD, where the rate is reported as bits per pixel (bpp) and is defined as:

$$R = \frac{B}{N \times W \times H} \quad (1)$$

where $B$ is the total amount of bits after compression by HEVC or NNC (depending on the strategy), $N$ is the number of rendered images, and $W$ and $H$ are the respective width and height in pixels. The distortion is reported as PSNR (for the *mugs* scene) and ST-LPIPS (for the *antique*, *train*, and *playground* scenes). These quality metrics are the most accurate for quality evaluation of NeRF-generated content in synthetic and real scenes, respectively, according to the large-scale subjective and objective study presented in [13]. An anchor was defined as the case where no compression was applied to the NN parameters or to the images used in the training process. This anchor serves as an upper-bound and cannot be applied in practice since it would correspond to lossless coding (of parameters or images) and thus a significant amount of rate would be spent.

### C. Experimental Results

The RD performance results are shown in Fig. 4, for the NN parameter-based and pixel-based streaming strategies and considering the selected visual scenes. As expected, all solutions obtained lower qualities than the respective anchors and converge to the quality provided by the anchor level as the rate increases. The following conclusions can also be taken:

*1) Pixel-based streaming strategy compression solution:* As expected, the use of HEVC Inter leads to a higher RD performance than HEVC Intra for all the scenes. This shows that HEVC Inter is capable of efficiently exploiting the correlation between the images selected for training (multiple perspectives of the visual scene).

*2) Streaming strategies comparison:* The pixel-based strategy outperforms the NN parameter-based strategy for the *mugs* and *antique* visual scenes. This result can be attributed to several factors. First, the HEVC codec used in the pixel-based streaming strategy is a rather efficient predictive codec with many prediction modes developed along several decades of research. Second, the NeRF learning process is robust to training images with quality degradation caused by quantization. In fact, the NeRF learning assumes that the scene represented by the training images is static and thus the quantization errors (which are typically uniformly distributed) are averaged out during training when multiple images provide a ray pointing to the same 3D position. This effect can be observed since the synthesized image often has higher perceptual quality than the training images.

*3) Impact of visual scene complexity:* The NN parameter-based streaming strategy performs better than the pixel-based strategy at higher rates for the *antique* (only for HEVC Intra), *train*, and *playground* scenes. Especially for the *train* and *playground* scenes, the set of training images is more difficult to compress with HEVC due to high spatial complexity. In particular, for those (360º real) scenes, it can be observed that the background content of the training images has more quantization noise at higher rates, resulting in an increased presence of distortions (e.g. floaters) for the pixel-based streaming strategy.

In conclusion, the NN parameter-based streaming strategy is clearly the best choice since it allows to reach the quality level of the anchor independently of the characteristics of the visual scene. Moreover, if the pixel-based strategy is used (e.g. due to the absence of a capable cloud/edge server), the selected training images can be compressed with lower quality if the visual scene is synthetic (or real but simple) while it is advisable to spend more rate to reach higher qualities for more complex 360º content.

## VI. Final Remarks

This paper provides a comprehensive RD performance evaluation between two NeRF streaming strategies with different performance-complexity tradeoffs and application use cases. As future work, it is suggested the design of compression techniques for NeRF models, aiming to enhance its performance which is especially relevant to broadcast and multicast applications.